\documentclass[preprintnumbers,twocolumn,prd,aps,superscriptaddress,footinbib,amsfonts,showpacs,floatfix]{revtex4-1}

\usepackage[T1]{fontenc}
\usepackage{blindtext}
\usepackage{amsmath}
\usepackage{amssymb}
\usepackage{graphics,graphicx}
\usepackage{dcolumn}
\usepackage{bm}
\usepackage{epsfig}
\usepackage[normalem]{ulem}
\usepackage[usenames]{color}
\usepackage{mathbbol}
\usepackage{epstopdf}
\usepackage{simplewick}
\usepackage[utf8]{inputenc} 
\usepackage{slashed}
\usepackage{soul}
\usepackage[dvipsnames]{xcolor}
\usepackage[export]{adjustbox}
\usepackage{mathtools}
\usepackage[colorlinks = true, citecolor = blue, linkcolor = blue, urlcolor = blue]{hyperref}

\allowdisplaybreaks

\begin{document}

\preprint{}

\title{\bf Simplified approach to extracting nucleon transversity in collinear \\ factorization using near-side energy-energy correlators}

\newcommand*{\UCLA}{Department of Physics and Astronomy, University of California, Los Angeles, CA 90095, USA}\affiliation{\UCLA}
\newcommand*{\Bhaumik}{Mani L. Bhaumik Institute for Theoretical Physics, University of California, Los Angeles, CA 90095, USA}\affiliation{\Bhaumik}
\newcommand*{\SB}{Center for Frontiers in Nuclear Science, Stony Brook University, Stony Brook, NY 11794, USA}\affiliation{\SB}
\newcommand*{\TU}{Department of Physics, SERC, Temple University, Philadelphia, Pennsylvania 19122, USA}\affiliation{\TU}
\newcommand*{\LVC}{Department of Physics, Lebanon Valley College, Annville, Pennsylvania 17003, USA}\affiliation{\LVC}

\author{Zhong-Bo Kang}\email{zkang@physics.ucla.edu}\affiliation{\UCLA}\affiliation{\Bhaumik}\affiliation{\SB}
\author{Andreas Metz}\email{metza@temple.edu}\affiliation{\TU}
\author{Daniel Pitonyak}\email{pitonyak@lvc.edu}\affiliation{\LVC}
\author{Congyue Zhang}\email{maxzhang2002@g.ucla.edu}\affiliation{\UCLA}\affiliation{\Bhaumik}

\begin{abstract}
\noindent We develop a novel strategy for accessing the transversity parton distribution function (PDF) of the nucleon within collinear factorization using near-side energy-energy correlators in the dihadron fragmentation framework.  We show how this removes the complications of previous approaches that must model either intrinsic parton transverse momentum  or resonances in the invariant mass distribution of a final-state dihadron.  We present leading-order analytical results for transverse-spin observables in semi-inclusive deep-inelastic scattering and electron-positron annihilation, highlighting their close similarity to the expressions one uses in extracting (un)polarized PDFs and (single-hadron) fragmentation functions in collinear factorization.  We make predictions for kinematics relevant for existing and future facilities that demonstrate the feasibility of an energy-energy correlator program in extracting the transversity PDF. 
\end{abstract}

\maketitle 

%%%%%%%%%%%%%%%%%%%%%%%%%%%%%%%%%%%%%
\section{Introduction and Motivation}
Energy-energy correlators (EECs), measured as a function of the opening angle $\chi$ between the two hadrons, have received intense interest over the last several years -- see an extensive review in Ref.~\cite{Moult:2025nhu}.  Experimental data on EECs in the $\chi\approx 0$ (near-side/collinear) and $\chi\approx \pi$ (away-side/back-to-back) regions show a clear transition from asymptotically free quarks and gluons to bound states of free hadrons~\cite{CELLO:1982rca,Fernandez:1984db,JADE:1984taa,PLUTO:1985yzc,Wood:1987uf,TASSO:1987mcs,TOPAZ:1989yod,ALEPH:1990vew,OPAL:1990reb,OPAL:1991uui,L3:1991qlf,L3:1992btq,SLD:1994idb,SLD:1994yoe,Bossi:2025xsi,CMS:2024mlf,ALICE:2024dfl,Bossi:2025nux}. The near-side free hadron and transition regions are ideally suited to be studied using dihadron fragmentation functions (DiFFs), with this relationship being formulated in Refs.~\cite{Lee:2025okn,Guo:2025zwb,Chang:2025kgq,Kang:2025zto,Guo:2025qnz}.  Connections of energy correlators to spin physics have also been pursued that have established pathways to extract the nucleon helicity $g_1(x)$~\cite{Gao:2025cwy} and transversity $h_1(x)$~\cite{Kang:2023big,Liu:2024kqt,Gao:2025evv,Cao:2025icu} parton distribution functions (PDFs), and transverse momentum dependent (TMD) functions, like the Sivers TMD PDF~\cite{Liu:2022wop,Kang:2023big,Liu:2024kqt,Bhattacharya:2025bqa} and Collins TMD fragmentation function~(FF)~\cite{Kang:2023big,Liu:2024kqt}.  

The transversity PDF~\cite{Ralston:1979ys} is highly important in its own right and has received particular attention since it can be used to compute the up and down quark tensor charges, $\delta u$ and $\delta d$, of the nucleon.  The nucleon tensor charges bridge different areas of nuclear physics.  The ability to measure certain beyond the Standard Model (BSM) couplings in free neutron beta decay (see, e.g., Refs.~\cite{Herczeg:2001vk,Erler:2004cx,Severijns:2006dr,Cirigliano:2013xha,Courtoy:2015haa,Gonzalez-Alonso:2018omy})
relies on knowledge of the isovector combination $(\delta u - \delta d)$.  In addition, $\delta u$ and $\delta d$ are ingredients for BSM physics in computing the nucleon electric dipole moment from those of the quarks (see, e.g., Refs.~\cite{Erler:2004cx,Pospelov:2005pr,Yamanaka:2017mef,Liu:2017olr}).  Other ways to determine the nucleon tensor charges include ab initio approaches in lattice QCD (LQCD)~\cite{Gupta:2018qil, Gupta:2018lvp, Yamanaka:2018uud, Hasan:2019noy, Alexandrou:2019brg, Harris:2019bih, Horkel:2020hpi, Alexandrou:2021oih, Park:2021ypf, Tsuji:2022ric, Bali:2023sdi, QCDSFUKQCDCSSM:2023qlx,Gao:2023ktu, Djukanovic:2024krw, Alexandrou:2024ozj, Wang:2025nsd} and model calculations~\cite{He:1994gz, Barone:1996un, Schweitzer:2001sr, Gamberg:2001qc, Pasquini:2005dk, Wakamatsu:2007nc, Lorce:2007fa, Yamanaka:2013zoa, Pitschmann:2014jxa, Xu:2015kta, Wang:2018kto, Liu:2019wzj}. 

Currently, a widely-used procedure in QCD phenomenology for extracting transversity relies on TMD factorization with single-hadron final states~\cite{Collins:1992kk, Anselmino:2007fs, Anselmino:2008jk, Anselmino:2013vqa, Anselmino:2015sxa, Kang:2015msa, Lin:2017stx, DAlesio:2020vtw, Cammarota:2020qcw, Gamberg:2022kdb,  Boglione:2024dal, Zeng:2024gun}. These reactions involve the transversity TMD $h_1(x,k_T)$, unpolarized TMD $f_1(x,k_T)$, Collins TMD FF $H_1^\perp(z,p_T)$, and the unpolarized TMD FF $D_1(z,p_T)$, whose dependence on the intrinsic transverse momentum $k_T$ and $p_T$ must be modeled, oftentimes through complicated functional forms~\cite{Kang:2015msa,Bacchetta:2017gcc,Bertone:2019nxa,Bacchetta:2019sam,Scimemi:2019cmh,Bacchetta:2022awv,Cerutti:2022lmb,Moos:2023yfa,Bacchetta:2024qre,Bacchetta:2025ara,Moos:2025sal,Rossi:2025pwh,Barry:2025glq,Kang:2026mod}.  

The other common approach is using transverse-spin observables with dihadron final states that can be analyzed in collinear factorization~\cite{Collins:1993kq,Jaffe:1997hf,Bianconi:1999cd,Radici:2001na,Bacchetta:2002ux,Bacchetta:2008wb,Bacchetta:2011ip,Bacchetta:2012ty,Radici:2015mwa,Radici:2016lam,Radici:2018iag,Cocuzza:2023oam,Cocuzza:2023vqs}.  In this case, the collinear PDFs $h_1(x)$ and $f_1(x)$ enter along with the DiFFs $H_1^{\sphericalangle}(z,M_h)$ and $D_1(z,M_h)$ ($M_h$ is the invariant mass of the dihadron). This strategy was initially conceived as a way to bypass the intricacies of TMD factorization~\cite{Bianconi:1999cd,Radici:2001na}. Nevertheless, with $H_1^{\sphericalangle}(z,M_h)$ and $D_1(z,M_h)$, one still deals with two-argument functions, and especially the $M_h$ dependence is complicated due to resonances, requiring ${\mathcal O}(100)$ parameters to model the DiFFs~\cite{Courtoy:2012ry,Cocuzza:2023vqs,Mahaut:2025hie}.  

In this article, we establish the framework for a novel strategy to extract the nucleon transversity PDF using near-side EECs in SIDIS and $e^+e^-$ annihilation. The observables couple $h_1(x)$ to EEC-DiFFs that have no dependence on $M_h$ and take on a simple functional form involving only a single variable.  Thus, the procedure for extracting $h_1(x)$ using EECs becomes essentially the same as that for $f_1(x)$ and $g_1(x)$.  We present numerical predictions that make a robust case that transverse-spin near-side EEC observables are measurable through a re-analysis of existing data from BELLE, COMPASS, and HERMES or future data from BESIII, Jefferson Lab, and the Electron-Ion Collider (EIC). 

%%%%%%%%%%%%%%%%%%%%%%%%%%%%%%%%%%%%%
\section{EEC Dihadron Fragmentation Functions}
We start with the operator definitions~\cite{Pitonyak:2023gjx} (see also Ref.~\cite{Pitonyak:2025lin}) of the two DiFFs, $D_1^{h_1h_2/q}$ and $H_1^{\sphericalangle\, h_1h_2/q}$, that will be relevant for our analytical results: 
\begin{align}
    &\!\!\!{\rm DiFF}(\xi_1,\xi_2,\vec{R}_T) = \!\frac{\xi^2}{64\pi^3\xi_1\xi_2}\!\int\! \!d^2\vec{k}_T\,\sum_X\hspace{-0.5cm}\int\! \int\!\!\frac{dx^+\!d^2\vec{x}_\perp}{(2\pi)^3}e^{ik\cdot x}\nonumber\\
    &\!\times {\rm Tr}\langle 0|\Gamma\,\psi_q(x)|P_1,P_2;X\rangle\langle P_1,P_2;X|\bar{\psi}_q(0)|0\rangle\big|_{x^-=0}\,,
\end{align}
where ${\rm DiFF}=D_1^{h_1h_2/q}$ when $\Gamma=\gamma^-$, and ${\rm DiFF}=-\frac{\epsilon_T^{ij}R_T^j}{M_h}H_1^{\sphericalangle\,h_1h_2/q}$ when $\Gamma=i\sigma^{i-}\gamma^5$.  The variables $\xi_1,\xi_2$ are the fractions of the light-cone momentum of the fragmenting quark $q$ (with momentum $k$) carried by the hadrons $h_1, h_2$, and $\vec{R}_T\equiv \frac{1}{2}(\vec{P}_{1T}-\vec{P}_{2T})$ is (half of) their relative transverse momentum. 
(We have suppressed gauge links, and a color average is understood.)  The total momentum of the dihadron is denoted by $P_h$, with $M_h^2=P_h^2$ its invariant mass, and $\xi\equiv \xi_1+\xi_2$.  The Levi-Civita tensor is defined as $\epsilon_T^{ij}\equiv \epsilon^{-+ij}$, with $\epsilon^{0123}=+1$.  We work in the ``dihadron frame'' where $P_h$ has no transverse component and a large lightcone-minus component.  

In Ref.~\cite{Kang:2025zto} (see also Refs.~\cite{Lee:2025okn,Guo:2025zwb,Chang:2025kgq,Guo:2025qnz}) we introduced the ``EEC-DiFF'' $\mathcal{D}^i(z_\chi, Q^2)$ ($i=q,g$), where $z_\chi\equiv \frac{1}{2}(1-\cos\chi)$ and $Q$ is the relevant hard scale, that was able to well describe data for near-side EECs in $e^+e^-$ annihilation. In calculating transverse-spin EEC observables involving dihadron final states, due to the fact that $H_1^{\sphericalangle\,h_1h_2/q}=-H_1^{\sphericalangle\,h_2h_1/q}$~\cite{Radici:2001na,Bacchetta:2006un,Cocuzza:2023vqs}, we find it becomes necessary to specify the species of the two hadrons being detected, rather than summing over $h_1,h_2$. We will also weight the relevant cross sections with factors of the hadron momentum fractions.  Therefore, we define the following non-perturbative functions for our analysis:
\begin{align}
\mathcal{D}^{h_1h_2/i}(z_\chi, Q^2)\equiv&\int \!\!d\xi_1 \!\int \!\!d\xi_2\int \!\!d^2\!\vec{R}_T\,\delta\!\left(\!z_\chi-\frac{R_T^2}{Q^2}\frac{\xi^2}{\xi_1^2\xi_2^2}\right)\nonumber\\
    &\times\, \xi_1\xi_2\,D_1^{h_1h_2/i}(\xi_1,\xi_2,\vec{R}_T)\,,\label{e:EEC_DiFF}\\[0.3cm]
\mathcal{H}^{h_1h_2/i}(z_\chi, Q^2)\equiv&\int \!\!d\xi_1 \!\int \!\!d\xi_2\int \!\!d^2\!\vec{R}_T\,\delta\!\left(\!z_\chi-\frac{R_T^2}{Q^2}\frac{\xi^2}{\xi_1^2\xi_2^2}\right)\nonumber\\
    &\hspace{-0.5cm}\times\, \xi_1\xi_2\,\,\frac{R_T}{M_h}H_1^{\sphericalangle\,h_1h_2/i}(\xi_1,\xi_2,\vec{R}_T)\,,\label{e:H1EEC_DiFF}
\end{align}
where $R_T\equiv|\vec{R}_T|$.  The evolution equation for $\mathcal{D}^{h_1h_2/i}(z_\chi, Q^2)$ was derived in Ref.~\cite{Kang:2025zto}.  The one for $\mathcal{H}^{h_1h_2/i}(z_\chi, Q^2)$ takes the same form, but the transversely polarized splitting kernels~\cite{Stratmann:2001pt} are used instead~\cite{Ceccopieri:2007ip,Pitonyak:2023gjx}.
\\

%%%%%%%%%%%%%%%%%%%%%%%%%%%%%%%%%%%%%
\section{Transverse-Spin Observables for Near-Side EECs}~We will study transverse-spin dependent near-side EECs for SIDIS, $e(l)\,N^\uparrow\!(P)\to\,e(l')\,(h_1(P_1)\, h_2(P_2))\, X$, and $e^+e^-$ annihilation to two almost back-to-back dihadrons, $e^+(l')e^-(l)\to (h_1(P_1)\,h_2(P_2))\,(\bar{h}_1(\bar{P}_1)\bar{h}_2(\bar{P}_2))\,X$, where the momenta of the particles involved are given in parentheses.   For both processes we introduce the variable $z_{12}\equiv \frac{1}{2}\!\left(\!1-\tfrac{\vec{P}_1\cdot\vec{P}_2}{|\vec{P}_1||\vec{P}_2|}\right) = \frac{1}{2}(1-\cos\theta_{12})$, where $\theta_{12}$ is the angle between $h_1$ and $h_2$. Up to power suppressed corrections $\sim 1/Q^2$, one can show~\cite{Lee:2025okn,Guo:2025zwb,Chang:2025kgq,Kang:2025zto,Guo:2025qnz}, 
\begin{equation}
    z_{12} = \frac{R_T^2}{Q^2}\frac{\tau^2}{\tau_1^2\tau_2^2}\,,\label{e:z12}
\end{equation}
with $\tau\equiv\tau_1+\tau_2$.  For SIDIS, $\tau_{1(2)}=P\cdot P_{1(2)}/P\cdot q$, with $q= l-l'$ and $Q^2= -q^2$, whereas for $e^+e^-$ annihilation, $\tau_{1(2)}=2P_{1(2)}\!\cdot q/Q^2=2E_{1(2)}/Q$, where $E_{1(2)}$ is the energy of hadron $h_{1(2)}$, and $Q^2= q^2$ is the squared center-of-mass (c.m.)~energy of the collision. The expression for $z_{12}$ in Eq.~\eqref{e:z12} specifically holds in the Breit frame for SIDIS and the c.m.~frame for $e^+e^-$ annihilation.

We define the transverse-spin EEC for SIDIS as
\begin{align}
    &{\rm EEC}_{\rm SIDIS}^{h_1h_2}
    \equiv \frac{d\Sigma_{\rm SIDIS}^{h_1h_2}}{dxdyd\phi_S dz_\chi d\phi}\nonumber\\
    &\;\;\equiv\int \!\!d\tau_1d\tau_2d^2\!\vec{R}_T\,\tau_1\tau_2\nonumber\\
    &\hspace{0.5cm}\times \frac{d\sigma_{\rm SIDIS}}{dxdyd\tau_1d\tau_2d^2\!\vec{R}_Td\phi_S}\,\delta(z_\chi-z_{12})\delta(\phi-\phi_{\!R_T})\,, \label{e:TEEC_SIDIS_def}
\end{align}
where an azimuthal angle dependence is included~\cite{Kang:2023big} to access spin effects, $x = Q^2/(2P\cdot q)$, $y=P\cdot q/P\cdot l$, and $\phi_S$ is the azimuthal angle of the nucleon's transverse spin vector.  For $e^+e^-$ annihilation we have
\begin{align}
    &{\rm EEC}_{e^+e^-}^{h_1h_2,\bar{h}_1\bar{h}_2} \equiv \frac{d\Sigma_{e^+e^-}^{h_1h_2,\bar{h}_1\bar{h}_2}}{dydz_\chi dz_{\bar{\chi}}d\phi d\bar{\phi}}\nonumber\\ 
    &\;\;\equiv \int \!\!d\tau_1d\tau_2d^2\!\vec{R}_T\int \!\!d\bar{\tau}_1d\bar{\tau}_2d^2\!\vec{\bar{R}}_T\,(\tau_1\tau_2)(\bar{\tau}_1\bar{\tau}_2)\nonumber\\
    &\hspace{0.5cm}\times \frac{d\sigma_{e^+e^-}}{ d\tau_1d\tau_2d^2\!\vec{R}_Td\bar{\tau}_1d\bar{\tau}_2d^2\!\vec{\bar{R}}_Tdy}\nonumber\\[0.1cm]
    &\hspace{0.5cm}\times \delta(z_\chi-z_{12})\delta(z_{\bar{\chi}}-z_{\overline{12}})\delta(\phi-\phi_{\!R_T})\delta(\bar{\phi}-\phi_{\!\bar{R}_T})\,, \label{e:TEEC_epem_def}
\end{align}
where $y=P_h\cdot l/P_h\cdot q$.  The leading-order (LO) results for the near-side dihadron contribution to ${\rm EEC}_{\rm SIDIS}^{h_1h_2}$ and ${\rm EEC}_{e^+e^-}^{h_1h_2,\bar{h}_1\bar{h}_2}$ can be found by following calculations of transverse-spin dependent dihadron cross sections~\cite{Boer:2003ya,Bacchetta:2008wb, Courtoy:2012ry,Radici:2015mwa,Matevosyan:2018icf, Artru:1995zu,Bacchetta:2002ux, Bacchetta:2003vn, Bacchetta:2011ip, Bacchetta:2012ty, Radici:2015mwa,Cocuzza:2023vqs}, although in this case one is differential in $(\tau_1,\tau_2,\vec{R}_T)$ instead of $(\tau,M_h)$.  The final expressions read, respectively,
\vspace{-0.5cm}
\begin{widetext}
\begin{align}
    {\rm EEC}_{\rm SIDIS}^{h_1h_2} &\overset{{\rm DiFF}}{=} \frac{\alpha_{em}^2}{\pi y Q^2}\sum_{q,\bar{q}}e_q^2\left[(1-y+\tfrac{y^2}{2})f_1^{q/N}(x)\,\mathcal{D}^{h_1h_2/q}(Z)-(1-y)\sin(\phi+\phi_S)h_1^{q/N}(x)\mathcal{H}^{h_1h_2/q}(Z)\right],\label{e:TEEC_SIDIS}\\
    {\rm EEC}_{e^+e^-}^{h_1h_2,\bar{h}_1\bar{h}_2}\!&\overset{{\rm DiFF}}{=}\frac{N_c\alpha_{em}^2}{\pi Q^2}\sum_{q,\bar{q}}e_q^2\left[(\tfrac{1}{2}-y+y^2)\mathcal{D}^{h_1h_2/q}(Z)\mathcal{D}^{\bar{h}_1\bar{h}_2/\bar{q}}(\bar{Z})+y(1-y)\cos(\phi+\bar{\phi})\mathcal{H}^{h_1h_2/q}(Z)\mathcal{H}^{\bar{h}_1\bar{h}_2/\bar{q}}(\bar{Z})\right],\label{e:TEEC_epem}
\end{align}
\end{widetext}
\vspace{-0.5cm}
where we have introduced the shorthand $Z\equiv z_\chi Q^2$ to emphasize that, due to the delta functions and integrations over $(\xi_1,\xi_2,\vec{R}_T)$ in Eqs.~\eqref{e:EEC_DiFF}, \eqref{e:H1EEC_DiFF}, $\mathcal{D}^{h_1,h_2/i}$ and $\mathcal{H}^{h_1,h_2/i}$ only depend on the {\it single variable} $z_\chi Q^2$ (up to an overall prefactor that depends just on $Q^2$).  We mention that for the DiFF contribution to the near-side EEC in $e^+e^-\to h_1h_2\,X$, the unpolarized result at next-to-leading order (NLO) was given in Ref.~\cite{Kang:2025zto}.  For the SIDIS case we present the NLO analytical expression in Supplemental Material~\cite{EEC_AUT_SM} as well as provide predictions for the EIC based on the EEC-DiFF $\mathcal{D}^i(Z)$ extracted in Ref.~\cite{Kang:2025zto}.  

The structure of Eqs.~\eqref{e:TEEC_SIDIS}, \eqref{e:TEEC_epem} is exactly the same as {\it collinear factorization} in SIDIS and $e^+e^-$ for {\it single-hadron} observables that are widely used in extractions of (un)polarized PDFs and FFs, where one encounters convolutions like $f_1(x)\otimes D_1(z)$ or $g_1(x)\otimes D_1(z)$. This highlights the advantage of our proposed program to extract the transversity PDF through these new observables.  We anticipate $\mathcal{D}^{h_1h_2/i}(Z)$ and $\mathcal{H}^{h_1h_2/i}(Z)$ will each take on a simple functional form.  In fact, in Ref.~\cite{Kang:2025zto}, we obtained successful phenomenology for near-side EECs in $e^+e^-$ annihilation using $\mathcal{D}^{i}(Z) = NQ^2\exp(-Z/\alpha)/(1+Z/\beta)$ ($N,\alpha,\beta$ were free parameters that in the present case may depend on flavor).

We note that factorization for the EEC dihadron observables \eqref{e:TEEC_SIDIS_def}, \eqref{e:TEEC_epem_def} relies on the factorization of the underlying dihadron cross sections.  As was discussed in Refs.~\cite{Rogers:2024nhb, Pitonyak:2025lin}, the same factorization theorems that have been proven for unpolarized single-hadron processes can be pushed through for those involving dihadron FFs.
While no NLO calculations for the transversely polarized cross sections exist yet, there have been such computations for related single-hadron processes $\ell p^\uparrow \to \ell \Lambda^\uparrow X$~\cite{Contogouris:1996xk} and $e^+e^-\to \Lambda^\uparrow \bar{\Lambda}^\uparrow$~(back-to-back)~$\!X$~\cite{Contogouris:1995xc}, which involve transversely polarized (anti)quarks in the initial and/or final states, just like in the present case.

\begin{figure*}[t]
    \centering
    \includegraphics[width=0.95\textwidth]{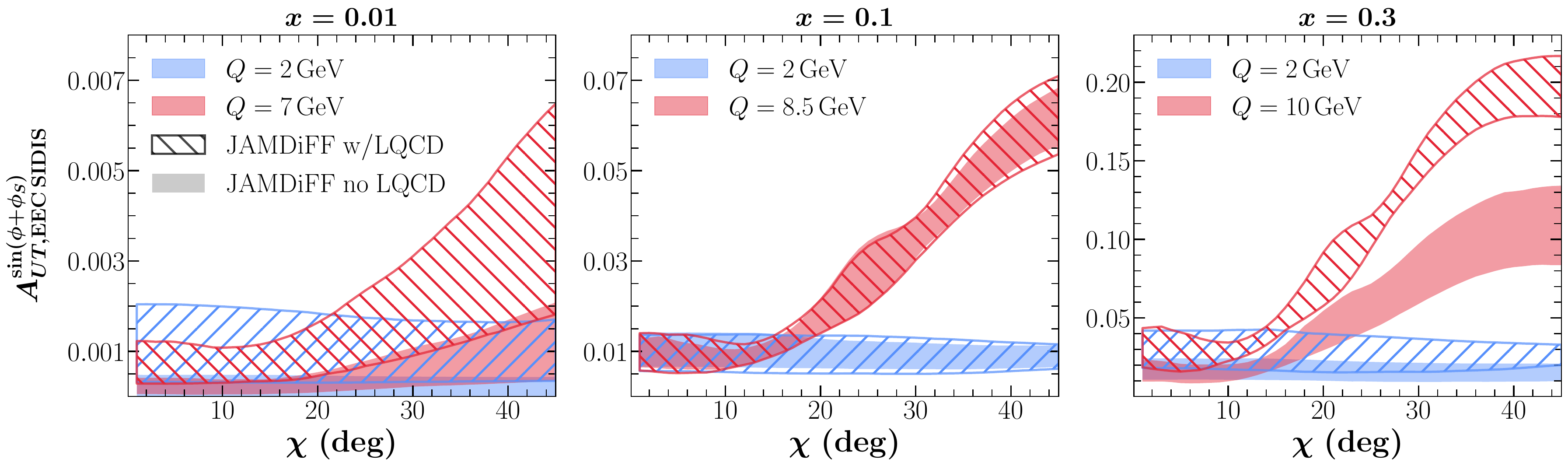}
    \caption{Predictions for $A_{UT,{\rm EEC\,SIDIS}}^{\sin(\phi+\phi_S)}$ vs.~$\chi$ at fixed $x=0.01,\,0.1,\,0.3$ (from left to right), each for two different values of $Q$. The filled bands show the central $68\%$ uncertainty, accounting for both the scan over $(a,b)$ parameter values and the statistical uncertainty from the non-perturbative JAMDiFF input.  Both the JAMDiFF with~(hatched) and without (solid) LQCD priors on the nucleon tensor charges are considered.}
    \label{fig:sidis-aut}
\end{figure*}

\begin{figure*}[t]
    \centering
    \includegraphics[width=0.95\textwidth]{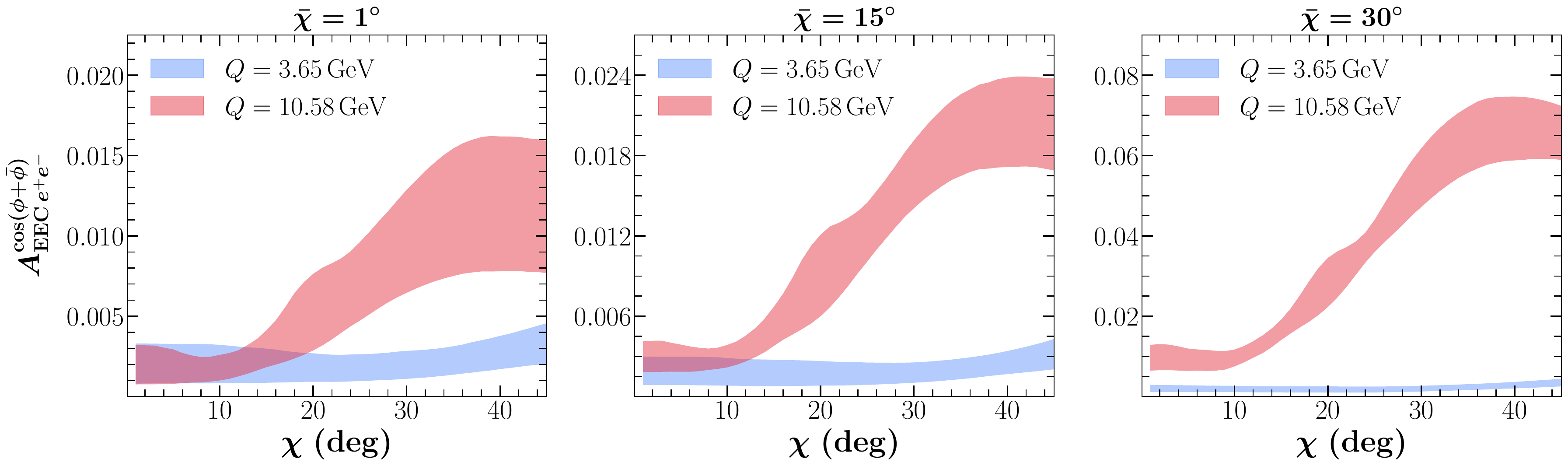}
    \caption{Predictions for $A_{{\rm EEC}\,e^+e^-}^{\cos(\phi+\bar{\phi})}$ vs.~$\chi$ at fixed $\bar{\chi}=1^\circ,\,15^\circ,\,30^\circ$ (from left to right) each for two different values of $Q$ for BESIII ($Q=3.65\,{\rm GeV}$) and BELLE ($Q=10.58\,{\rm GeV}$). 
    Both JAMDiFF with and without LQCD priors give nearly identical results, so only the latter is shown.}
    \label{fig:ee-aut}
\end{figure*}

We mention that other ways have been proposed and pursued previously in which the transversity PDF could be extracted through collinear factorization without invoking dihadron final states.  For example, $A_{TT}$ in doubly transversely polarized proton-proton collisions for hadron, jet, or lepton-pair (Drell-Yan) probes $h_1(x)\otimes h_1(x)$~\cite{Ralston:1979ys,Hidaka:1978pi,Artru:1989zv,  Ji:1992ev,Jaffe:1996ik,Soffer:2002tf}. In addition, the transverse spin-transfer observable $D_{TT}$ in $p^\uparrow p\to \Lambda^\uparrow X$ or $\ell\,N^\uparrow\to \ell'\Lambda^\uparrow X$ couples $h_1(x)$ to the transversity FF $H_1(z)$~\cite{deFlorian:1998am,Anselmino:2000ga,Xu:2004es,Xu:2005ru,Kang:2021kpt}.  Some measurements exist for $A_{TT}$ and $D_{TT}$~\cite{STAR:2012hth,STAR:2018fqv,COMPASS:2021bws,STAR:2023hwu}, but all of them are compatible with zero.  On the other hand, based on our predictions in Figs.~\ref{fig:sidis-aut},~\ref{fig:ee-aut} (discussed more later), not only should the observables proposed below in Eqs.~\eqref{e:AUTSIDIS}, \eqref{e:AUTepem} be sizable, but results for them could become available through a re-analysis of existing dihadron data from BELLE, HERMES, and COMPASS~\cite{Belle:2011cur,HERMES:2008mcr,Belle:2017rwm,COMPASS:2023cgk,COMPASS:2025dwu} or future measurements at BESIII~\cite{BESIII:2020nme}, Jefferson Lab (CLAS~\cite{Burkert:2020akg} and SoLID~\cite{JeffersonLabSoLID:2022iod}), and the EIC~\cite{AbdulKhalek:2021gbh}.

From the EECs given in Eqs.~\eqref{e:TEEC_SIDIS}, \eqref{e:TEEC_epem}, focusing on $\pi^+\pi^-$ final states, we define the following asymmetries as the ratio of the azimuthal-dependent to the azimuthal-independent terms:
\begin{align}
    A_{UT,{\rm EEC\,SIDIS}}^{\sin(\phi+\phi_S)} &\equiv -\frac{\sum_{q,\bar{q}}e_q^2\,h_1^{q/N}(x)\mathcal{H}^{\pi^+\pi^-/q}(Z)}{\sum_{q,\bar{q}}e_q^2\,f_1^{q/N}(x)\,\mathcal{D}^{\pi^+\pi^-/q}(Z)}\,,\label{e:AUTSIDIS} \\[0.1cm]
    A_{{\rm EEC}\,e^+e^-}^{\cos(\phi+\bar{\phi)}} &\equiv \frac{\sum_{q,\bar{q}}e_q^2\,\mathcal{H}^{\pi^+\pi^-/q}(Z)\mathcal{H}^{\pi^+\pi^-/\bar{q}}(\bar{Z})}{\sum_{q,\bar{q}}e_q^2\,\mathcal{D}^{\pi^+\pi^-/q}(Z)\mathcal{D}^{\pi^+\pi^-/\bar{q}}(\bar{Z})}\,.\label{e:AUTepem}
\end{align}
Note that these definitions do not contain depolarization ($y$-dependent) factors.  The asymmetries \eqref{e:AUTSIDIS}, \eqref{e:AUTepem} can be at most equal to 1 in magnitude. In addition, information is needed for the EEC in $e^+e^-\to h_1h_2 X$ (without the sum over $h_1,h_2$): 
\begin{align}
    &{\rm EEC}^{\pi^+\pi^-}_{e^+e^-} \equiv \frac{1}{\sigma_t}\frac{d\Sigma^{\pi^+\pi^-}_{e^+e^-}}{d\chi}\nonumber\\
    &\;\;\equiv\frac{\sin\chi}{2}\frac{1}{\sigma_t}\int \!\!d\tau_1d\tau_2d^2\vec{R}_T\,\tau_1\tau_2 \frac{d\sigma^{e^+e^-}}{ d\tau_1d\tau_2d^2\!\vec{R}_T}\delta(z_\chi-z_{12})\,\nonumber\\
    &\;\;\overset{{\rm DiFF}}{=}\frac{\sin\chi}{2}\frac{1}{\sigma_t}\frac{4\pi N_c\alpha_{em}^2}{3Q^2} \sum_{q,\bar{q}}e_q^2\,\mathcal{D}^{\pi^+\pi^-/q}(Z)\,,\label{e:EECpi}
\end{align}
where $\sigma_t$ is the total cross section for $e^+e^-\to {\rm hadrons}$.  The last line  in Eq.~\eqref{e:EECpi} is the LO result.  (We have defined ${\rm EEC}^{\pi^+\pi^-}_{e^+e^-}$ to be differential in $\chi$ instead of $z_\chi$ and normalized by $\sigma_t$ in order to be in full analogy with the ``standard'' EEC in $e^+e^-\to h_1h_2 X$.) One can then simultaneously extract $h_1(x), \, \mathcal{D}^{\pi^+\pi^-/i}(Z), \, \mathcal{H}^{\pi^+\pi^-/i}(Z)$ from (future) data on $A_{UT,{\rm EEC\,SIDIS}}^{\sin(\phi+\phi_S)}, \, A_{{\rm EEC}\,e^+e^-}^{\cos(\phi+\bar{\phi)}}, \, {\rm EEC}^{\pi^+\pi^-}_{e^+e^-}$, analogous to the approach of Refs.~\cite{Cocuzza:2023oam,Cocuzza:2023vqs} but with the very helpful simplification of not having to model the $M_h$ dependence of DiFFs.

%%%%%%%%%%%%%%%%%%%%%%%%%%%%%%%%%%%%%
\section{Modeling of the EEC DiFFs} In order to motivate measurements of Eqs.~\eqref{e:AUTSIDIS}, \eqref{e:AUTepem}, and \eqref{e:EECpi}, we will give numerical predictions for kinematics relevant for existing and future facilities in the next section.  We first must model the EEC-DiFFs $\mathcal{D}^{\pi^+\pi^-/i}(Z)$ and $\mathcal{H}^{\pi^+\pi^-/i}(Z)$, and we do so by relating them to the DiFFs $D_1^{\pi^+\pi^-/i}(\xi,M_h)$ and $H_1^{\sphericalangle\,\pi^+\pi^-/i}(\xi,M_h)$, respectively, that have been extracted~\cite{Cocuzza:2023oam,Cocuzza:2023vqs} from current dihadron measurements in SIDIS, $e^+e^-$ annihilation, and proton-proton collisions~\cite{Belle:2011cur,HERMES:2008mcr,Belle:2017rwm,COMPASS:2023cgk,STAR:2015jkc,STAR:2017wsi}.  However, we emphasize that this approach is only for the purpose of making predictions for this study.  Once experimental data on Eqs.~\eqref{e:AUTSIDIS}, \eqref{e:AUTepem}, and \eqref{e:EECpi} are available, one would parameterize $\mathcal{D}^{h_1h_2/i}(Z)$ and $\mathcal{H}^{h_1h_2/i}(Z)$ using a functional form like $NQ^2\exp(-Z/\alpha)/(1+Z/\beta)$ mentioned above instead of connecting them to $M_h$-dependent DiFFs.

As we derive in Supplemental Material~\cite{EEC_AUT_SM}, one has 
\begin{align}
    \mathcal{D}^{h_1h_2/i}(Z) = &\frac{Q^2}{32}\int_0^1 \!d\xi\int_{-1}^1\!d\zeta\,\frac{\xi^4(1-\zeta^2)^2}{M_h}\nonumber\\
    &\times D_1^{h_1h_2/i}(\xi,\zeta,M_h)\bigg |_{M_h=\widetilde{M}_h}\,,\label{e:DZDMh}
\end{align}
where $\widetilde{M}_h=\sqrt{Z\xi^2(1-\zeta^2)/4+M_{h,min}^2}$, with $M_{h,min}^2=2M_1^2/(1+\zeta)+2M_2^2/(1-\zeta)$ the lower bound on $M_h^2$~\cite{Bianconi:1999uc,Bacchetta:2002ux,Radici:2001na}, and $\zeta\equiv (\xi_1-\xi_2)/\xi$. The exact same relation~\eqref{e:DZDMh} holds between $\mathcal{H}^{h_1h_2/i}(Z)$ and $H_1^{\sphericalangle\, h_1h_2/i}(\xi,\zeta,M_h)$.  The DiFFs that have been extracted from experiment, though, are only functions of $(\xi,M_h)$~\cite{Courtoy:2012ry,Cocuzza:2023oam,Cocuzza:2023vqs,Mahaut:2025hie}.  Therefore, we expand the $\zeta$ dependence of $D_1^{h_1h_2/i}(\xi,\zeta,M_h)$ and $H_1^{\sphericalangle\, h_1h_2/i}(\xi,\zeta,M_h)$ in terms of Legendre polynomials up to second order, similar to the partial wave expansion of DiFFs formulated in Ref.~\cite{Bacchetta:2002ux}.  Each coefficient is assumed to be proportional to the $(\xi,M_h)$-dependent DiFF.  Specifically,
\begin{align}
    D_1(\xi,\zeta,M_h)&\approx D_1(\xi,M_h)\!\left[\frac{1}{2}+a\zeta+\frac{1}{2}b(3\zeta^2-1)\!\right]\!,\label{e:D1Leg}\\[0.1cm]
     H_1^{\sphericalangle}(\xi,\zeta,M_h)&\approx H_1^{\sphericalangle}(\xi,M_h)\!\left[\frac{1}{2}+\tilde{a}\zeta+\frac{1}{2}\tilde{b}(3\zeta^2-1)\!\right],\label{e:H1Leg}
\end{align}
where $a,b,\tilde{a},\tilde{b}$ are unknown constants, and we have dropped the $h_1h_2/i$ superscript on the functions for brevity.  Note that this parameterization of the $\zeta$ dependence automatically satisfies $\int_{-1}^1 d\zeta\, D_1(\xi,\zeta,M_h)=D_1(\xi,M_h)$, as must be the case from the number density interpretation of the DiFFs~\cite{Pitonyak:2023gjx}.  The positivity bounds on $ D_1(\xi,\zeta,M_h)$ and $H_1^{\sphericalangle}(\xi,\zeta,M_h)$, namely, $D_1(\xi,\zeta,M_h)\ge 0$ and $|H_1^{\sphericalangle}(\xi,\zeta,M_h)|\le D_1(\xi,\zeta,M_h)$~\cite{Bacchetta:2002ux}, also allow one to place constraints on the possible values of $a,b,\tilde{a},\tilde{b}$.  Since the positivity bounds also apply to the $(\xi,M_h)$-dependent DiFFs, the $D_1(\xi,\zeta,M_h)\ge 0$ bound requires $-\frac{1}{2}\le b\le 1, \, |a|\le \frac{1}{2}+b$, and, when $b>0$ and $|a|\le 3b$, $|a|\le \sqrt{3b(1-b)}$ (see Supplemental Material~\cite{EEC_AUT_SM} for more details).  The $|H_1^{\sphericalangle}(\xi,\zeta,M_h)|\le D_1(\xi,\zeta,M_h)$ bound is most naturally satisfied when $\tilde{a}=a$ and $\tilde{b}=b$. 

%%%%%%%%%%%%%%%%%%%%%%%%%%%%%%%%%%%%%
\section {Numerical Predictions} We now give numerical predictions at LO for $A_{UT,{\rm EEC\,SIDIS}}^{\sin(\phi+\phi_S)}, \, A_{{\rm EEC}\,e^+e^-}^{\cos(\phi+\bar{\phi)}}, \, {\rm EEC}^{\pi^+ \pi^-}_{e^+e^-}$ in Eqs.~\eqref{e:AUTSIDIS}, \eqref{e:AUTepem}, and \eqref{e:EECpi} utilizing Eqs.~\eqref{e:DZDMh}, \eqref{e:D1Leg}, and \eqref{e:H1Leg}.  We use the JAMDiFF analysis~\cite{Cocuzza:2023vqs} as input for $h_1(x), D_1^{\pi^+\pi^-/i}(\xi,M_h)$, $H_1^{\sphericalangle\,\pi^+\pi^-/i}(\xi,M_h)$, which are evolved to $\mathcal{O}(\alpha_s)$.  The evolution of $\mathcal{D}^{h_1h_2/i}(Z)$ and $\mathcal{H}^{h_1h_2/i}(Z)$ is ``inherited'' from the DGLAP evolution of their corresponding DiFFs via~\eqref{e:DZDMh}.
We construct the central $68\%$ uncertainty bands from an ensemble of $1000$ Monte Carlo predictions. For each prediction, we independently draw, with replacement, one replica from JAMDiFF for $\{h_1(x), D_1^{\pi^+\pi^-/i}(\xi,M_h)$, $H_1^{\sphericalangle\,\pi^+\pi^-/i}(\xi,M_h)\}$, one replica from CT18NLO~\cite{Hou:2019qau,Hou:2016sho} for $f_1(x)$, and one of the $200$ uniformly distributed points in the $(a,b)$ allowed region (see Supplemental Material~\cite{EEC_AUT_SM}) and evaluate the result for the three scale choices $\mu \in \{Q/2,Q,2Q\}$. We note that the scale variation for any given prediction causes an increase in the average size of the uncertainty band of typically around $5$-$10\%$ or less and no more than $20\%$.
We also consider separately the JAMDiFF extraction with and without including priors from LQCD on the nucleon tensor charges.

In Fig.~\ref{fig:sidis-aut}, we present $A_{UT,{\rm EEC\,SIDIS}}^{\sin(\phi+\phi_S)}$ vs.~$\chi$ for fixed values of $x=0.01,0.1,0.3$ and two different values of $Q$ for each, representative of the kinematics of potential future measurements at the EIC. The upper limit on the angle $\chi$ is chosen so that $\sqrt{Z} =\sqrt{z_\chi}\,Q< 4.5~{\rm GeV}$, which allows the DiFF in Eq.~(12) as one integrates to stay mostly within the $(\xi,M_h)$ range where the JAMDiFF analysis (whose DiFFs we use as input) is valid.
We find for larger $Q$ the asymmetry generally increases with $\chi$ and with $x$. The effect is largest at $x=0.3$ due to $h_1(x)$ peaking in that region, becoming as large as $20\%$ in the JAMDiFF-with-LQCD scenario at $Q=10\,{\rm GeV}$. Note that for $x=0.3$, COMPASS has $Q\sim 5\,{\rm GeV}$, so $A_{UT,{\rm EEC\,SIDIS}}^{\sin(\phi+\phi_S)}$ there falls in between the two curves in the right panel of Fig.~\ref{fig:sidis-aut} (for a given JAMDiFF scenario), while for CLAS and SoLID, $Q\sim 2\,{\rm GeV}$ at $x=0.3$, which is given by the blue curve.  For HERMES, at $x=0.15$, $Q\sim 2\,{\rm GeV}$, so $A_{UT,{\rm EEC\,SIDIS}}^{\sin(\phi+\phi_S)}$ there would basically follow the blue curve in the central panel of Fig.~\ref{fig:sidis-aut}. 

In Fig.~\ref{fig:ee-aut}, we display $A_{{\rm EEC}\,e^+e^-}^{\cos(\phi+\bar{\phi})}$ vs.~$\chi$ for BELLE and BESIII c.m.~energies at different fixed values of $\bar{\chi}=1^\circ,\,15^\circ,\,30^\circ$.  We find at the largest $\bar{\chi}$, the asymmetry is close to $8\%$ at BELLE and below $1\%$ for BESIII.  We note that for the asymmetries $A_{UT,{\rm EEC\,SIDIS}}^{\sin(\phi+\phi_S)}$ and $A_{{\rm EEC}\,e^+e^-}^{\cos(\phi+\bar{\phi)}}$, the contribution to the uncertainty from the scanning over the allowed region of $(a,b)$ values was negligible compared to the (statistical) theoretical uncertainties of the non-perturbative JAMDiFF functions.

\begin{figure}[t]
    \centering
    \includegraphics[width=0.9\columnwidth]{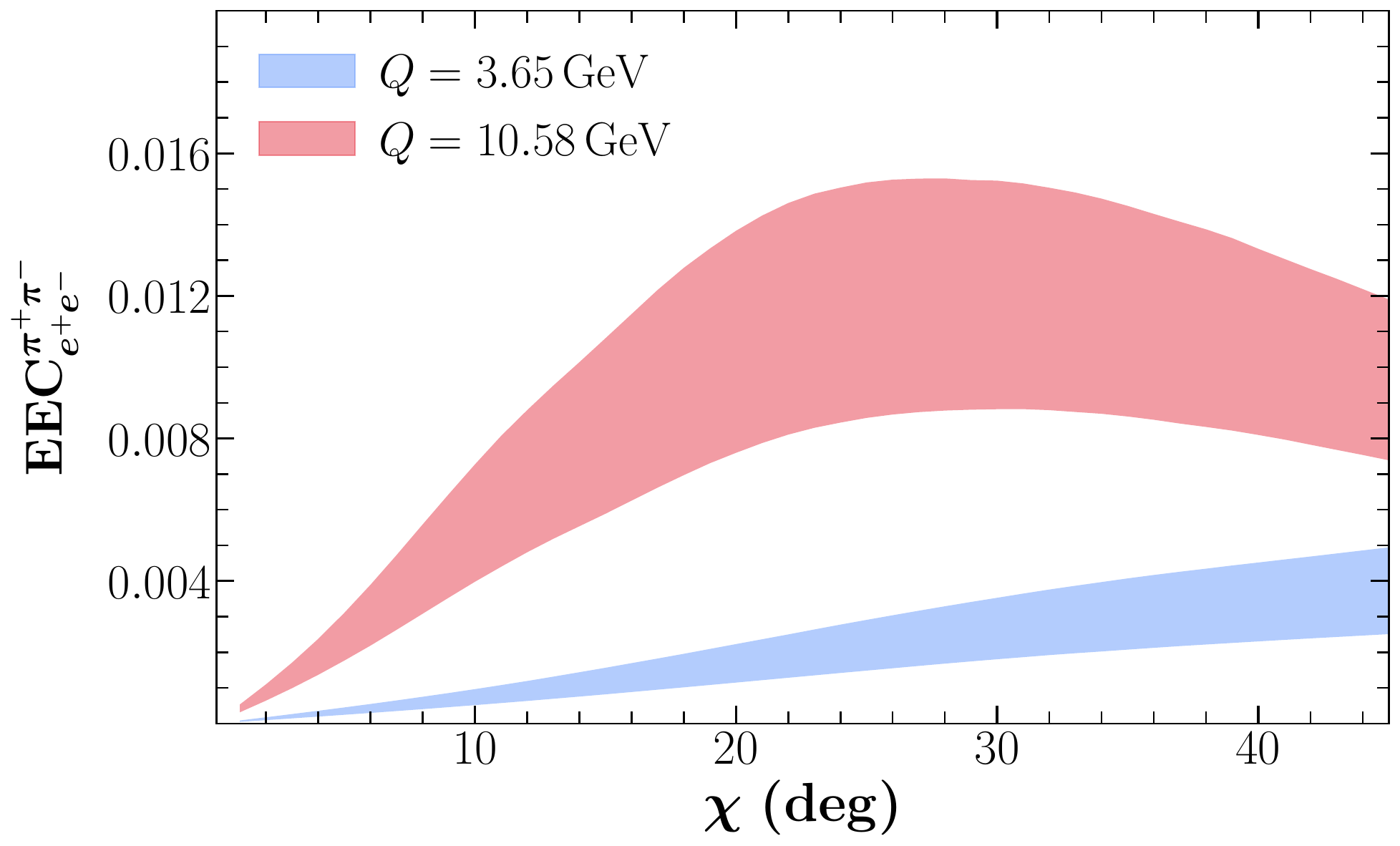}
    \caption{Predictions for ${\rm EEC}^{\pi^+ \pi^-}_{e^+e^-}$ vs.~$\chi$ at BESIII ($Q=3.65\,{\rm GeV}$) and BELLE ($Q=10.58\,{\rm GeV}$) kinematics.
    \vspace{-0.5cm}}
    \label{fig:ee-eec}
\end{figure}

In Fig.~\ref{fig:ee-eec}, we plot ${\rm EEC}^{\pi^+ \pi^-}_{e^+e^-}$ vs.~$\chi$ for BELLE and BESIII c.m.~energies.  We find that for $Q=10.58\,{\rm GeV}$, ${\rm EEC}^{\pi^+ \pi^-}_{e^+e^-}$ has a similar shape (peak and falloff) to what one would expect given existing lower-energy ``standard'' EEC measurements at $Q=14\,{\rm GeV}, 22\,{\rm GeV}$~\cite{TASSO:1987mcs,JADE:1984taa}. For $Q=3.65\,{\rm GeV}$ the energy is too low to see that behavior (one would have to extend to larger angles that fall outside of the region where the JAMDiFF analysis is valid). 
Notably, the BELLE ${\rm EEC}^{\pi^+ \pi^-}_{e^+e^-}$ prediction in
Fig.~\ref{fig:ee-eec} peaks at
$Q\sin(\chi/2)\sim 2.6 \,{\rm GeV}$, close to the $2.8 \, {\rm GeV}$ quark/gluon to free hadron transition peak found in the near-side EEC analyses of
Refs.~\cite{Liu:2024lxy, Kang:2025zto, Herrmann:2025fqy}. Since the ${\rm EEC}^{\pi^+ \pi^-}_{e^+e^-}$
prediction is generated from JAMDiFF DiFFs through Eq.~\eqref{e:DZDMh}, without using EEC data as input, this agreement suggests that the transition scale is a robust feature of the underlying hadronization dynamics. We also highlight that there are no resonance structures in ${\rm EEC}^{\pi^+ \pi^-}_{e^+e^-}$,  unlike for the cross section measurement $d\sigma/dzdM_h$ from BELLE for $\pi^+\pi^-$ production~\cite{Belle:2017rwm}, which causes the aforementioned complicated modeling of $D_1^{\pi^+\pi^-/i}(\xi,M_h)$ in the current DiFF approach to extracting $h_1(x)$.  Unlike for the asymmetries, the uncertainty band for  ${\rm EEC}^{\pi^+ \pi^-}_{e^+e^-}$ vs.~$\chi$ in Fig.~\ref{fig:ee-eec} is almost entirely due to scanning over the allowed region of $(a,b)$ values.

%%%%%%%%%%%%%%%%%%%%%%%%%%%%%%%%%%%%%
\section{Conclusions and Outlook}
We have established a new paradigm for extracting the nucleon transversity PDF by analyzing transverse-spin asymmetries for near-side (collinear) EECs in SIDIS and $e^+e^-$ annihilation.  Our numerical predictions make a robust case for measuring these observables through a re-analysis of existing data from BELLE, HERMES, and COMPASS or future data from BESIII, Jefferson Lab, and the EIC. This will open up a new avenue for a simplified extraction of $h_1(x)$ compared to the current TMD and dihadron approaches, where the procedure now essentially becomes analogous to canonical extractions of $f_1(x), g_1(x)$, and $D_1(z)$ in collinear factorization. \\

{\it Note Added}:~While finalizing this manuscript, a preprint on a STAR measurement of EECs for dihadron-in-jet appeared~\cite{STAR:2026epw}. Our  approach can be extended to this proton-proton case, which we leave for future work.

\section*{Acknowledgments}
We thank A.~Vossen for a helpful discussion on dihadron experimental measurements.  This work was supported by the National Science Foundation under Grants  No.~PHY-2515057~(Z.K., C.Z.), No.~PHY-2412792~(A.M.), and No.~PHY-2308567, PHY-2608477~(D.P.). 

%merlin.mbs apsrev4-1.bst 2010-07-25 4.21a (PWD, AO, DPC) hacked
%Control: key (0)
%Control: author (8) initials jnrlst
%Control: editor formatted (1) identically to author
%Control: production of article title (-1) disabled
%Control: page (0) single
%Control: year (1) truncated
%Control: production of eprint (0) enabled
%

%%%%%%%%%%%%%%%%%%%%%%%%%%%%%%%%%%%%%
\clearpage
\onecolumngrid
\section{Supplemental Material}
\setcounter{equation}{0}
\setcounter{figure}{0}
% \pagenumbering{arabic}
% \renewcommand{\thepage}{S\arabic{page}}  
\renewcommand{\thesubsection}{S\arabic{subsection}}   
\setcounter{secnumdepth}{2}
\renewcommand{\thetable}{S\arabic{table}}   
\renewcommand{\thefigure}{S\arabic{figure}}
\renewcommand{\theequation}{S\arabic{equation}}

%%%%%%%%%%%%%%%%%%%%%%%%%%%%%%%%%%%%%
\subsection*{Unpolarized EEC in Semi-Inclusive DIS:~Analytical Result and Numerical Predictions}

\begin{figure*}[t]
    \centering
    \includegraphics[width=0.45\textwidth]{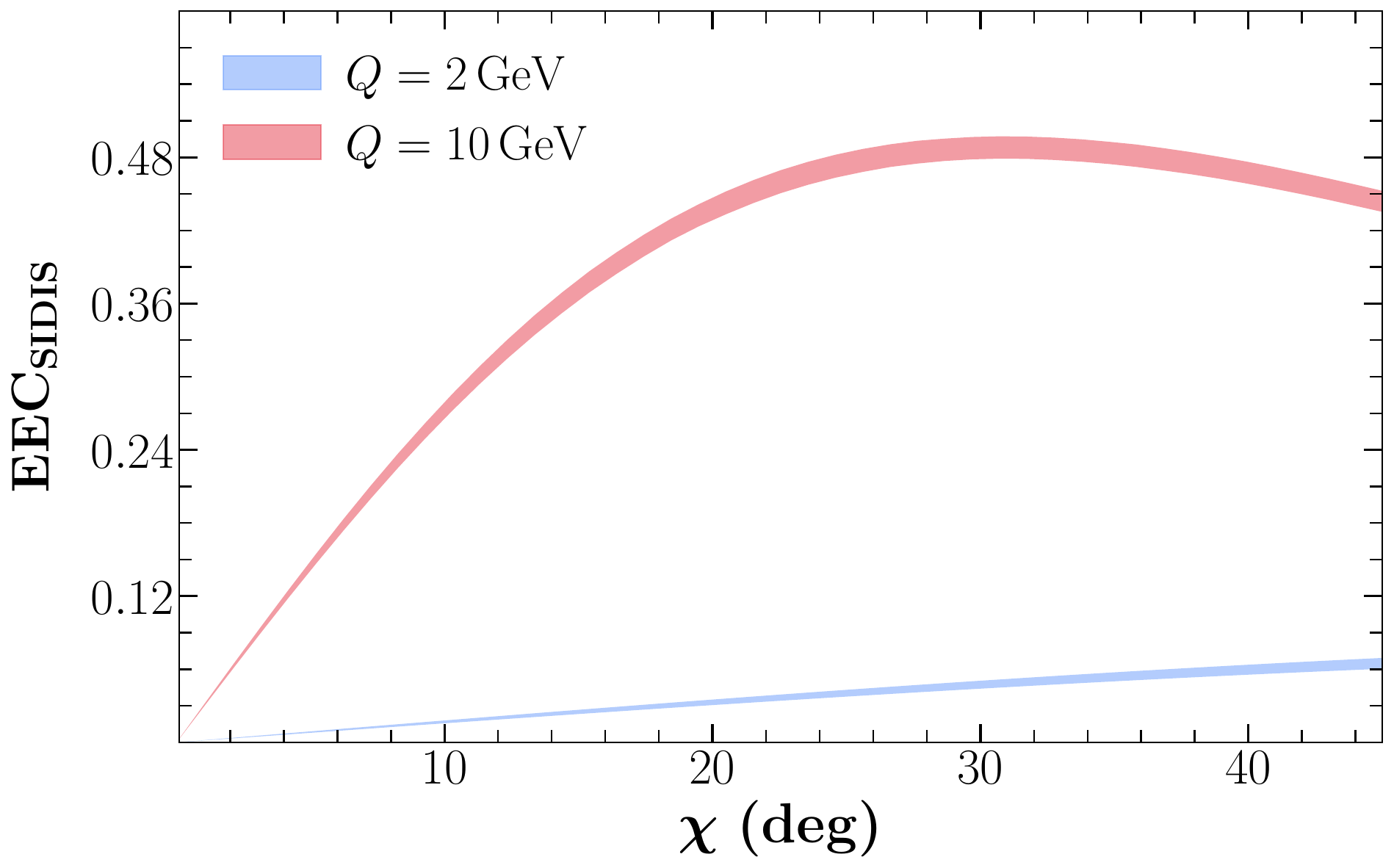}
    \caption{Predictions for ${\rm EEC_{SIDIS}}$ vs.~$\chi$ for two different values of $Q$ representative of EIC kinematics.  Note that at LO, and assuming a flavor-independent $\mathcal{D}^q(Z)$, the $x$ dependence of the observable cancels out.}
    \label{fig:sidis-eec}
\end{figure*}

From Refs.~\cite{Rogers:2024nhb, Pitonyak:2025lin} we know that the terms in the cross section $(d\sigma/dxdyd\tau_1 d\tau_2 d^2\!\vec{R}_T)^{eN\to e'h_1h_2X}$ that only involve dihadron fragmentation will have exactly the same structure as $(d\sigma/dxdyd\tau)^{eN\to e'hX}$ with $D_1^{h/i}\to D_1^{h_1h_2/i}$. 
However, the replacement $d\beta/\beta\to d\beta/\beta^2$ is also needed in the integration measure associated with the DiFFs -- see, e.g., Eq.~(16b) of Ref.~\cite{Pitonyak:2025lin}. Notably, the partonic cross sections are the same in both cases.  Based on the known factorization formula for $(d\sigma/dxdyd\tau)^{eN\to e'hX}$~\cite{deFlorian:1997zj}, one can then immediately write down 
\begin{equation}
    \frac{d\sigma^{eN\to e'h_1h_2X}}{dxdyd\tau_1 d\tau_2 d^2\!\vec{R}_T} \overset{\rm DiFF}{=} \frac{2\pi\alpha_{em}^2}{Q^2}\left[\frac{(1+(1-y)^2)}{y}\,2F_1(x,\tau_1,\tau_2,\vec{R}_T)+\frac{2(1-y)}{y}\,F_L(x,\tau_1,\tau_2,\vec{R}_T)\right],\,
\end{equation}
where
\begin{align}
    &2F_1(x,\tau_1,\tau_2,\vec{R}_T) = \sum_{q,\bar{q}}e_q^2\Bigg\{f_1^{q/N}(x)\,D_1^{h_1h_2/q}(\tau_1,\tau_2,\vec{R}_T)\nonumber\\[0.1cm]
    &\hspace{1.5cm}+\,\frac{\alpha_s}{2\pi}\int_x^1\!\frac{d\alpha}{\alpha}\int_\tau^1\!\frac{d\beta}{\beta^2}\bigg[f_1^{q/N}(\tfrac{x}{\alpha})\,\mathbb{C}^1_{qq}(\alpha,\beta)\,D_1^{h_1h_2/q}(\tfrac{\tau_1}{\beta},\tfrac{\tau_2}{\beta},\vec{R}_T)
    + f_1^{q/N}(\tfrac{x}{\alpha})\,\mathbb{C}^1_{gq}(\alpha,\beta)\,D_1^{h_1h_2/g}(\tfrac{\tau_1}{\beta},\tfrac{\tau_2}{\beta},\vec{R}_T)\nonumber\\
    &\hspace{4.5cm}+ f_1^{g/N}(\tfrac{x}{\alpha})\,\mathbb{C}^1_{qg}(\alpha,\beta)\,D_1^{h_1h_2/q}(\tfrac{\tau_1}{\beta},\tfrac{\tau_2}{\beta},\vec{R}_T)\bigg]\Bigg\},\\[0.3cm]
    %%%%%%%%%%%%%%%%
    &F_L(x,\tau_1,\tau_2,\vec{R}_T)=\frac{\alpha_s}{2\pi}\sum_{q,\bar{q}}e_q^2\int_x^1\!\frac{d\alpha}{\alpha}\int_\tau^1\!\frac{d\beta}{\beta^2}\bigg[f_1^{q/N}(\tfrac{x}{\alpha})\,\mathbb{C}^L_{qq}(\alpha,\beta)\,D_1^{h_1h_2/q}(\tfrac{\tau_1}{\beta},\tfrac{\tau_2}{\beta},\vec{R}_T)
    \nonumber\\
    &\hspace{3.25cm}+ f_1^{q/N}(\tfrac{x}{\alpha})\,\mathbb{C}^L_{gq}(\alpha,\beta)\,D_1^{h_1h_2/g}(\tfrac{\tau_1}{\beta},\tfrac{\tau_2}{\beta},\vec{R}_T)+ f_1^{g/N}(\tfrac{x}{\alpha})\,\mathbb{C}^L_{qg}(\alpha,\beta)\,D_1^{h_1h_2/q}(\tfrac{\tau_1}{\beta},\tfrac{\tau_2}{\beta},\vec{R}_T)\bigg],
\end{align}
with $\mathbb{C}^{1,L}_{ij}$ being the NLO coefficient functions that can be found in Appendix C of Ref.~\cite{deFlorian:1997zj}.  From here we can calculate a ``standard'' EEC for SIDIS, defined as 
\begin{align}
    {\rm EEC}_{\rm SIDIS} 
    &\equiv \frac{d\Sigma_{\rm SIDIS}}{dxdyd\chi}\frac{1}{\frac{d\sigma_{\rm DIS}}{dxdy}}\equiv \frac{1}{\frac{d\sigma_{\rm DIS}}{dxdy}}\frac{\sin\chi}{2}\sum_{h_1,h_2}\int \!\!d\tau_1d\tau_2d^2\!\vec{R}_T\,\tau_1\tau_2\, \frac{d\sigma^{eN\to e'h_1h_2X}}{dxdyd\tau_1d\tau_2d^2\!\vec{R}_T}\,\delta(z_\chi-z_{12})\nonumber\\
    &\overset{\rm DiFF}{=} \frac{1}{\frac{d\sigma_{\rm DIS}}{dxdy}} \frac{\sin\chi}{2}\!\left(\frac{2\pi\alpha_{em}^2}{Q^2}\!\right)\!\!\left[\frac{(1+(1-y)^2)}{y}\,2\mathcal{F}_1(x,Z)+\frac{2(1-y)}{y}\,\mathcal{F}_L(x,Z)\right], \label{e:EECSIDIS_sum}
\end{align}
where $d\sigma_{\rm DIS}/dxdy$ is the inclusive DIS cross section, and
\begin{align}
    &2\mathcal{F}_1(x,Z) = \sum_{q,\bar{q}}e_q^2\Bigg\{f_1^{q/N}(x)\,\mathcal{D}^{q}(Z)\nonumber+\,\frac{\alpha_s}{2\pi}\int_x^1\!\frac{d\alpha}{\alpha}\int_\tau^1\!d\beta\,\beta^2\bigg[f_1^{q/N}(\tfrac{x}{\alpha})\,\mathbb{C}^1_{qq}(\alpha,\beta)\,\mathcal{D}^q(\beta^2 Z)
    + f_1^{q/N}(\tfrac{x}{\alpha})\,\mathbb{C}^1_{gq}(\alpha,\beta)\,\mathcal{D}^g(\beta^2 Z)\nonumber\\
    &\hspace{8cm}+ f_1^{g/N}(\tfrac{x}{\alpha})\,\mathbb{C}^1_{qg}(\alpha,\beta)\,\mathcal{D}^q(\beta^2 Z)\bigg]\Bigg\},\\[0.3cm]
    %%%%%%%%%%%%%%%%
    &\mathcal{F}_L(x,Z)=\frac{\alpha_s}{2\pi}\sum_{q,\bar{q}}e_q^2\int_x^1\!\frac{d\alpha}{\alpha}\int_\tau^1\!d\beta\,\beta^2\bigg[f_1^{q/N}(\tfrac{x}{\alpha})\,\mathbb{C}^L_{qq}(\alpha,\beta)\,\mathcal{D}^q(\beta^2 Z)
    + f_1^{q/N}(\tfrac{x}{\alpha})\,\mathbb{C}^L_{gq}(\alpha,\beta)\,\mathcal{D}^g(\beta^2 Z)\nonumber\\
    &\hspace{5cm}+ f_1^{g/N}(\tfrac{x}{\alpha})\,\mathbb{C}^L_{qg}(\alpha,\beta)\mathcal{D}^q(\beta^2 Z)\bigg].
\end{align}
Here, $\mathcal{D}^q(Z), \mathcal{D}^g(Z)$ are the EEC-DiFFs extracted in Ref.~\cite{Kang:2025zto} from the EEC in $e^+e^-\to h_1h_2 X$, and we understand $\mathcal{D}^i(\beta^2 Z)$ as making the replacement $Q^2\to \beta^2 Q^2$ in Eqs.~\eqref{e:EEC_DiFF}, \eqref{e:H1EEC_DiFF}. We then can make a LO prediction for ${\rm EEC_{SIDIS}}$ using $\mathcal{D}^q(Z), \mathcal{D}^g(Z)$ from Ref.~\cite{Kang:2025zto} as input, which is shown in Fig.~\ref{fig:sidis-eec}. Note that at LO the DIS cross section will cancel out a corresponding factor in $d\Sigma_{\rm SIDIS}/dxdyd\chi$ since $\mathcal{D}^q(Z)$ from Ref.~\cite{Kang:2025zto} is assumed to be flavor independent. Such a cancellation will not occur at higher orders. 

%%%%%%%%%%%%%%%%%%%%%%%%%%%%%%%%%%%%%
\subsection*{\boldmath More Details on the Modeling of the EEC DiFFs $\mathcal{D}^{\pi^+\pi^-/i}(Z)$ and $\mathcal{H}^{\pi^+\pi^-/i}(Z)$}

Starting from Eq.~\eqref{e:EEC_DiFF} in the main text, we can use the relation~\cite{Pitonyak:2023gjx} 
\begin{equation}
    D_1^{h_1h_2/i}(\xi_1,\xi_2,\vec{R}_T) = \frac{4}{\pi\xi M_h(1-\zeta^2)}\,D_1^{h_1h_2/i}(\xi,\zeta,M_h)\,,
\end{equation}
along with the variable transformation $d\xi_1d\xi_2d^2\!\vec{R}_T=\frac{\xi(1-\zeta^2)}{16}d\xi d\zeta d\phi_{R_T}dM_h^2$, and the fact that
\begin{equation}
    \frac{R_T^2}{Q^2}\frac{\xi^2}{\xi_1^2\xi_2^2} = \frac{4}{Q^2\xi^2(1-\zeta^2)}(M_h^2-M_{h,min}^2)\,,
\end{equation}
where $M_{h,min}^2=2M_1^2/(1+\zeta)+2M_2^2/(1-\zeta)$ and $\zeta\equiv (\xi_1-\xi_2)/\xi$, to write
\begin{equation}
    \mathcal{D}^{h_1h_2/i}(Z) = \frac{Q^2}{32}\int_0^1 \!d\xi\int_{-1}^1\!d\zeta\,\frac{\xi^4(1-\zeta^2)^2}{M_h} D_1^{h_1h_2/i}(\xi,\zeta,M_h)\bigg |_{M_h=\widetilde{M}_h}\,,
\end{equation}
with $\widetilde{M}_h=\sqrt{Z\xi^2(1-\zeta^2)/4+M_{h,min}^2}$.  This is Eq.~\eqref{e:DZDMh} in the main text.  An analogous derivation leads to the exact same relation between $\mathcal{H}^{h_1h_2/i}(Z)$ and $H_1^{\sphericalangle\,h_1 h_2/i}(\xi,\zeta,M_h)$.

\begin{figure}[h!]
    \centering
    \includegraphics[width=0.65\columnwidth]{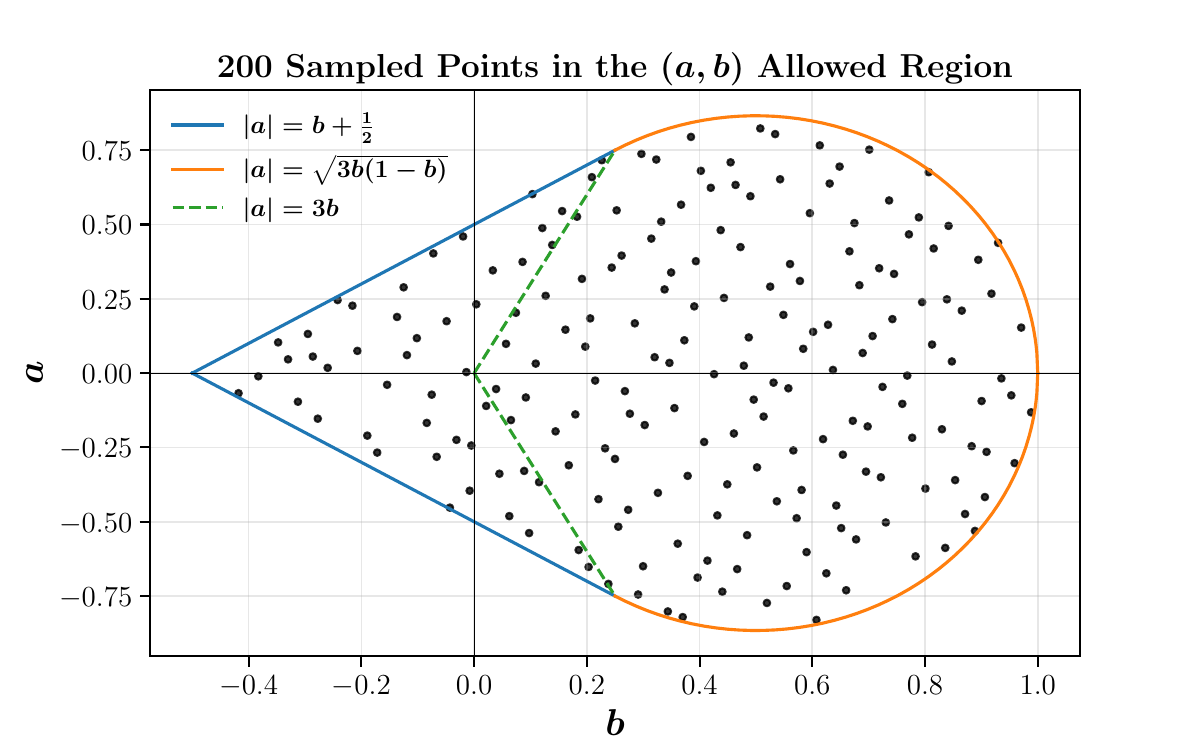}
    \caption{Allowed region for the $(a,b)$ parameters in Eq.~\eqref{e:D1Leg} due to the positivity bound on $D_1(\xi,\zeta,M_h)$.  The black points show the 200 $(a,b)$ samples used for our numerical predictions.  We also set $\tilde{a}=a$ and $\tilde{b}=b$ in Eq.~\eqref{e:H1Leg} to satisfy the $|H_1^{\sphericalangle}(\xi,\zeta,M_h)|\le D_1(\xi,\zeta,M_h)$ bound.} 
    \label{fig:ab}
\end{figure}

Now we turn to the Legendre expansions in Eqs.~\eqref{e:D1Leg}, \eqref{e:H1Leg} and provide more details on the region of $(a,b)$ explored in generating our numerical predictions.  Since $D_1(\xi,M_h)\ge 0$, we must have $f(\zeta)=\frac{1}{2}+a\zeta+\frac{1}{2}b(3\zeta^2-1)\ge 0$ $\forall \zeta \in [-1,1]$ in order for $D_1(\xi,\zeta,M_h)\ge 0$ to also be satisfied.  The extremum of $f(\zeta)$ is at $\zeta_0=-\frac{a}{3b}$.  If $b>0$, then $f(\zeta)$ is a concave up parabola and $\zeta_0$ gives a minimum.  If $|a|\le 3b$, then $\zeta_0\in[-1,1]$, and it must be required that $f(\zeta_0)\ge 0$, which implies $|a|\le \sqrt{3b(1-b)}$.  Otherwise, it is the value of $f(\zeta)$ at the endpoints $\zeta=-1$ and $\zeta=1$ that must be $\ge 0$, giving us the condition $|a|\le b+\frac{1}{2}$, which also implies $b\ge -\frac{1}{2}$. At $\zeta=0$, $f(\zeta)\ge 0$ also restricts $b\le 1$, and, therefore, $-\frac{1}{2}\le b \le 1$.  If $b<0$, then $f(\zeta)$ is a concave down parabola.  Even if $\zeta_0 \in [-1,1]$, $f(\zeta_0)$ is a maximum, so what matters is only that $f(\zeta=-1)$ and $f(\zeta=1)$ are both $\ge 0$, and we revert back to the condition $|a|\le b+\frac{1}{2}$.  Therefore, putting all of this together we have
\begin{align}
|a|\le
\begin{cases}
     b+\frac{1}{2}\;{\rm if}\; -\frac{1}{2}\le b\le 0\\
    b+\frac{1}{2} \;{\rm if}\;0<b\le 1\;{\rm and}\; |a|>3b\\
    \sqrt{3b(1-b)}\;{\rm if}\;0<b\le 1\;{\rm and}\; |a|\le 3b\,.
\end{cases}
\end{align}
Figure~\ref{fig:ab} shows the allowed $(a,b)$ region along with the points (sampled to uniformly cover the area) used in our numerical predictions of $A_{UT,{\rm EEC\,SIDIS}}^{\sin(\phi+\phi_S)}, \, A_{{\rm EEC}\,e^+e^-}^{\cos(\phi+\bar{\phi)}}, \, {\rm EEC}^{\pi^+ \pi^-}_{e^+e^-}$  (Figs.~\ref{fig:sidis-aut}, \ref{fig:ee-aut}, \ref{fig:ee-eec} in the main text), as a way to quantify the uncertainty due to the choice of $(a,b)$.  The $|H_1^{\sphericalangle}(\xi,\zeta,M_h)|\le D_1(\xi,\zeta,M_h)$ bound is most naturally satisfied when $\tilde{a}=a$ and $\tilde{b}=b$.
Lastly, we mention the term proportional to $\zeta$ in Eqs.~\eqref{e:D1Leg}, \eqref{e:H1Leg} gives a component to the integrand in Eq.~\eqref{e:DZDMh} that is odd in $\zeta$ and, consequently, it does not contribute directly to the EEC-DiFFs. 
Nevertheless, we retain that term in the Legendre expansion because it is part of the full positivity analysis:~the allowed two-dimensional $(a,b)$ region constrains the $b$-dependent contribution to the EEC-DiFFs and, therefore, enters the uncertainty estimate.

\end{document}